\documentclass[%
 reprint,
 superscriptaddress,
 amsmath,amssymb,
 aps,
 prapplied
]{revtex4-2}

\usepackage{xcolor}
\usepackage{graphicx}
\usepackage{dcolumn}
\usepackage{bm}
\usepackage{hyperref}

\begin{document}

\preprint{APS/123-QED}

\title{Micrometer-sized magnonic three-port rectilinear circulator}
\author{Kevin S. Weber}
\email{kevin.weber@imt-atlantique.fr}
\affiliation{Département Micro-Ondes, Institut Mines-Télécom (IMT) Atlantique, Technopole Brest-Iroise CS83818,
CEDEX 03, 29238 Brest, France}
\affiliation{Laboratoire des Sciences et Techniques de l’Information de la Communication et de la Connaissance
(Lab-STICC), Unité Mixte de Recherche (UMR) 6285, CNRS, Technopole Brest-Iroise CS83818, CEDEX 03,
29238 Brest, France}
\affiliation{QZabre AG, Neunbrunnenstrasse 50, 8050 Zürich, Switzerland}

\author{Loic Temdie}
\affiliation{Département Micro-Ondes, Institut Mines-Télécom (IMT) Atlantique, Technopole Brest-Iroise CS83818,
CEDEX 03, 29238 Brest, France}
\affiliation{Laboratoire des Sciences et Techniques de l’Information de la Communication et de la Connaissance
(Lab-STICC), Unité Mixte de Recherche (UMR) 6285, CNRS, Technopole Brest-Iroise CS83818, CEDEX 03,
29238 Brest, France}
\affiliation{SPEC, CEA, CNRS, Université Paris-Saclay, CEA Saclay, 91191 Gif-sur-Yvette Cedex, France}

\author{Vincent Castel}
\affiliation{Département Micro-Ondes, Institut Mines-Télécom (IMT) Atlantique, Technopole Brest-Iroise CS83818,
CEDEX 03, 29238 Brest, France}
\affiliation{Laboratoire des Sciences et Techniques de l’Information de la Communication et de la Connaissance
(Lab-STICC), Unité Mixte de Recherche (UMR) 6285, CNRS, Technopole Brest-Iroise CS83818, CEDEX 03,
29238 Brest, France}

\author{Timmy Reimann}
\affiliation{INNOVENT e.V. Technologieentwicklung, Pruessingstrasse 27B, 07745 Jena, Germany}

\author{Morris Lindner}
\affiliation{INNOVENT e.V. Technologieentwicklung, Pruessingstrasse 27B, 07745 Jena, Germany}

\author{Carsten Dubs}
\affiliation{INNOVENT e.V. Technologieentwicklung, Pruessingstrasse 27B, 07745 Jena, Germany}

\author{Yves Henry}
\affiliation{Unité Mixte de Recherche (UMR) 7504, CNRS, Institut de Physique et Chimie des Matériaux de Strasbourg,
Université de Strasbourg (IPCMS), CEDEX, 67000 Strasbourg, France}

\author{Matthieu Bailleul}
\affiliation{Unité Mixte de Recherche (UMR) 7504, CNRS, Institut de Physique et Chimie des Matériaux de Strasbourg,
Université de Strasbourg (IPCMS), CEDEX, 67000 Strasbourg, France}

\author{Vincent Vlaminck}
\email{vincent.vlaminck@imt-atlantique.fr}
\affiliation{Département Micro-Ondes, Institut Mines-Télécom (IMT) Atlantique, Technopole Brest-Iroise CS83818,
CEDEX 03, 29238 Brest, France}
\affiliation{Laboratoire des Sciences et Techniques de l’Information de la Communication et de la Connaissance
(Lab-STICC), Unité Mixte de Recherche (UMR) 6285, CNRS, Technopole Brest-Iroise CS83818, CEDEX 03,
29238 Brest, France}

\date{18 November 2026}

\begin{abstract}
     The development of miniaturized non-reciprocal microwave technologies compatible with integrated circuit architectures remains a critical challenge for modern information technology. Here, we present the first experimental characterization of a micron-sized prototypical magnon circulator. Taking advantage of the chiral excitation of spin-waves via nanowire gratings, we propose an original design of a circulator involving three channels of rectilinear and unidirectional spin-wave beams. We demonstrate via a full 3-port spin-wave spectroscopy a genuine spin-wave circulation between the three ports. The narrow frequency band of operation can be tuned over a broad range of frequencies ($2$--$8$\,GHz) with both an external field of up to $100$\,mT, and the dimensions of the grating specifying the wavevectors. This proposed scheme opens up possibilities for architectures of integrated and miniaturized non-reciprocal microwave devices.

\end{abstract}

\maketitle

    Non-reciprocal components such as circulators and isolators are indispensable for protecting microwave systems from reflections, isolating radar signals, and shielding qubits in quantum architectures\cite{Harris2012ModernFerrites}. However, these functionalities are not easily achieved with standard electric-field devices. The current ferrite-based technologies relying on the gyrotropic nature of magnetization remain bulky and off-chip, which limits their miniaturization into microscale electronics. This necessitates the development of alternative non-reciprocal technologies compatible with 5G/6G standards and integrated circuit technology \cite{Lira2012ElectricallyChip, Reiskarimian2016Magnetic-freeCommutation, Mahoney2017On-chipCirculator, Ranzani2019CirculatorsApproaches}. 
    Spin-waves offer a promising platform to develop a class of non-reciprocal devices compatible with integrated circuit technologies thanks to various intrinsic forms of non-reciprocities in frequency \cite{Grassi2020Slow-Wave-basedDiode, Haidar2014, Gladii2016}, and amplitude \cite{Schneider2008PhaseAntenna,Chen2019,Temdie2023HighBeams,Temdie2023ChiralGrating,Devolder2023,Wagle2026ShapingBeams}. Besides, magnonics enables a scalability down to sub-$100$\,nm sizes\cite{Sluka2019,Heinz2020,Temdie2024,Krma2025, Kruglyak2021ChiralMagnonics, Otalora2016Curvature-InducedDispersion}, as well as a tunability in a broad frequency range of the microwave spectrum, making it a field of choice for miniaturizing microwave components \cite{Dieny2020,Barman2021TheRoadmap}. So far, only theoretical efforts have been carried out to study the feasibility of a magnonic circulator, based primarily on numerical simulations of complex architectures. Among these studies, the inverse-design approach requires a magnetic landscape made of submicron squared pixels \cite{Wang2021Inverse-designDevices}, while another simulation study showing unidirectional coupling schemes relies on complex multilayered geometries including ports on both sides of the film \cite{Szulc2020Spin-waveCoupling, Lan2015Spin-WaveDiode}. While these simulations open routes for miniaturizing circulators, they require challenging nanoscale dimensions \cite{Zhao2023Three-terminalCirculator}, which poses significant challenges for both experimental realization and threshold sensitivity. Consequently, a practical micron-scale experimental demonstration of a circulator remains a critical missing link for integrated magnonic circuits. 
    In this study, we experimentally demonstrate spin-wave circulation at the micron scale on a device based on rectilinear unidirectional spin-wave beams propagating in a thin YIG film. We present results on two different prototypes showcasing genuine spin-wave circulation between three ports, also allowing multi-band operation by engineering the wavevector spectrum directly from the transducer dimensions. By combining this architecture with the frequency selectivity, tunability, and power dependence of spin-waves, we propose a scheme for the miniaturization of circulators.\\

\begin{figure}[!htbp]
\includegraphics[width=\columnwidth]{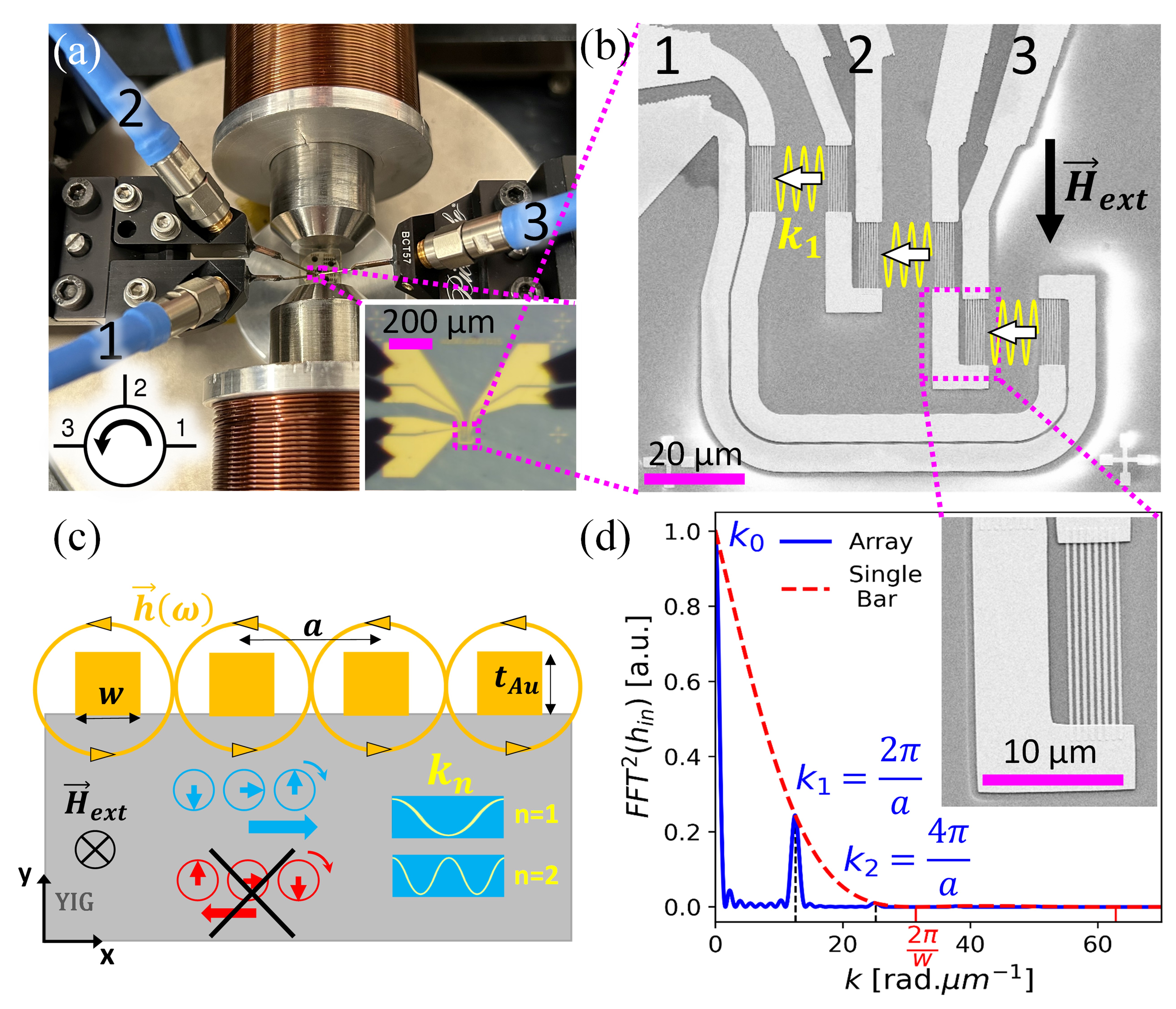}
\caption{\label{fig:1 Setup, SEM, CHirality, FFT}
            (a) 3-port spin-wave spectroscopy setup. (inset) Optical microscope image of the circulator prototype.
            (b) SEM image of the circulator. Arrows indicate the rectilinear spin-wave channels and circulation direction defined by the external field polarity.
            (c) Schematic of the chiral phase-matching mechanism between the microwave field distribution of the gold nanowires (yellow circles) and the spin-wave phase profiles (blue and red circles).
            (d) Fourier transform of the excitation field from the grating, showing a comb of wavevector $k_n=n\cdot 2\pi/a$. (inset) 
            High-magnification SEM of the gold nano-bar transducers.}
\end{figure}

\begin{figure}[!tbp]
    \includegraphics[width=\columnwidth]{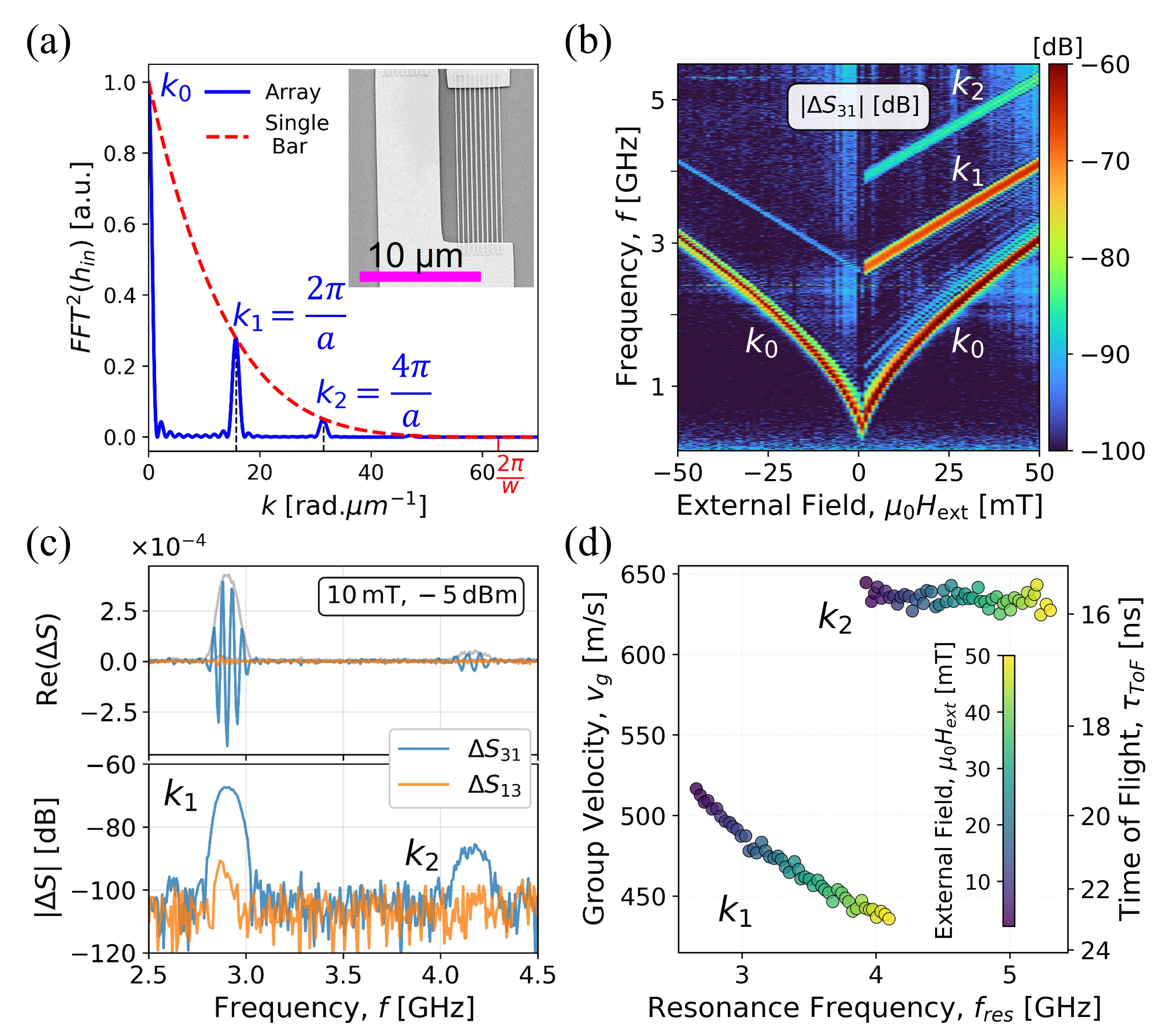}
    \caption{
        Second device with reduced nano-bar width $w=100$\,nm and periodicity $a=400$\,nm.
        (a) Fourier transform of the excitation with an inset displaying an SEM image of the reduced nano-bar array. 
        (b) Frequency-field map of $\Delta S_{31}$ (acquired using $\mu_0 H_{\text{ref}} = 250$\,mT and $-5$\,dBm input power) revealing a second chiral wavevector branch $k_2$.
        (c) Corresponding spectra at $10$\,mT showing unidirectional transmission for both $k_1$ and $k_2$ peaks, (top) real part $\text{Re}(S_{31})$, and (bottom) magnitude $\lvert S_{31}\rvert$.
        (d) Field dependence of the group velocity $v_g$ and the time of flight $\tau_{\text{ToF}}$ for both $k_1$ and $k_2$ wavevector peaks (error bars are omitted for clarity as they generally fall within the marker size).
    }
    \label{fig:4 Device 2}
\end{figure}

    The spin-wave circulator device shown in Fig.~\ref{fig:1 Setup, SEM, CHirality, FFT}-(a,b) consists of three pairs of gold nano-bar gratings fabricated on top of a $55$\,nm-thick Yttrium Iron Garnet (YIG) film grown via liquid phase epitaxy \cite{Dubs2020}. This design functions as a three rectilinear beam channel architecture covering a total working area of around $100 \times 100$\,$\mu\text{m}^2$. Each grating acts as a chiral transducer, allowing for unidirectional transmission of spin-waves between two ports \cite{Temdie2023ChiralGrating}. The circulation scheme is achieved by translating the pairs of gratings, in order to create three separate channels, through which spin-wave beams propagate rectilinearly from one port to another.\\
    The essence of the non-reciprocity used in our device is sketched in Fig.~\ref{fig:1 Setup, SEM, CHirality, FFT}-(c). It relies on the chiral excitation of spin-waves in the Damon-Eschbach configuration, which originates from the matching of the distribution of the microwave Oersted fields generated by the nanowire gratings (yellow circles) with the spin-wave phase profile sketched with blue and red circles respectively for right and left propagating waves. As illustrated in Fig.~\ref{fig:1 Setup, SEM, CHirality, FFT}-(c), only the waves propagating to the right-hand side of the static magnetization direction $\vec{M}$ will effectively be excited, accordingly with the cross-product rule \cite{Kostylev2013,Yu2021ChiralRadiation,Wagle2026ShapingBeams}:
        \begin{equation}
            \frac{\vec{k}}{\lVert\vec{k}\rVert} \times \frac{\vec{M}}{M_s} = \vec{u_y}
            \label{eq:RightHandRule}
        \end{equation}
    where $\vec{u_y}$ is the unitary vector pointing out-of-the plane in the direction from the film to the grating. In other words, this unidirectional excitation, and thus the direction of circulation, can be inverted with the polarity of the external bias field $\vec{H}_{ext}$, or also if the grating is located below the YIG film.\\
    Furthermore, the grating's geometry provides a spectral excitation signature in the form of a wavevector comb corresponding to integral values of the spatial periodicity $a$ of the nanowires \cite{Temdie2023ChiralGrating,Yu2013OmnidirectionalCoupler}, namely $k_n = n \cdot \frac{2\pi}{a}, n\in\mathbb{N}_0$, as shown in Fig.~\ref{fig:1 Setup, SEM, CHirality, FFT}-(d) with the Fourier transform of the excitation field.
    This allows for a specific selection of several narrow frequency bands, which can be adjusted with the grating's dimensions. Both investigated devices share a gold thickness of $t_{\text{Au}}=40$\,nm, a nano-bar length of $13$\,$\mu$m, and array grating width of $4$\,$\mu$m. The first device shown in the scanning electron microscopy (SEM) image in the inset of Fig.~\ref{fig:1 Setup, SEM, CHirality, FFT}-(d) utilizes a nano-bar width $w=200$\,nm, and a periodicity $a=500$\,nm with an inter-transducer distance $d=15$\,$\mu$m. The second device, with its corresponding nano-bar displayed in the inset of Fig.~\ref{fig:4 Device 2}-(a), features reduced dimensions with $w=100$\,nm, $a=400$\,nm, and a distance of $d=10$\,$\mu$m. 
    In the following, we characterize the performance of each device demonstrating the circulation over multi-bands of frequency by engineering the wavevector spectrum with the gratings dimensions.

\begin{figure*}[!htbp]
    \includegraphics[width=\textwidth]{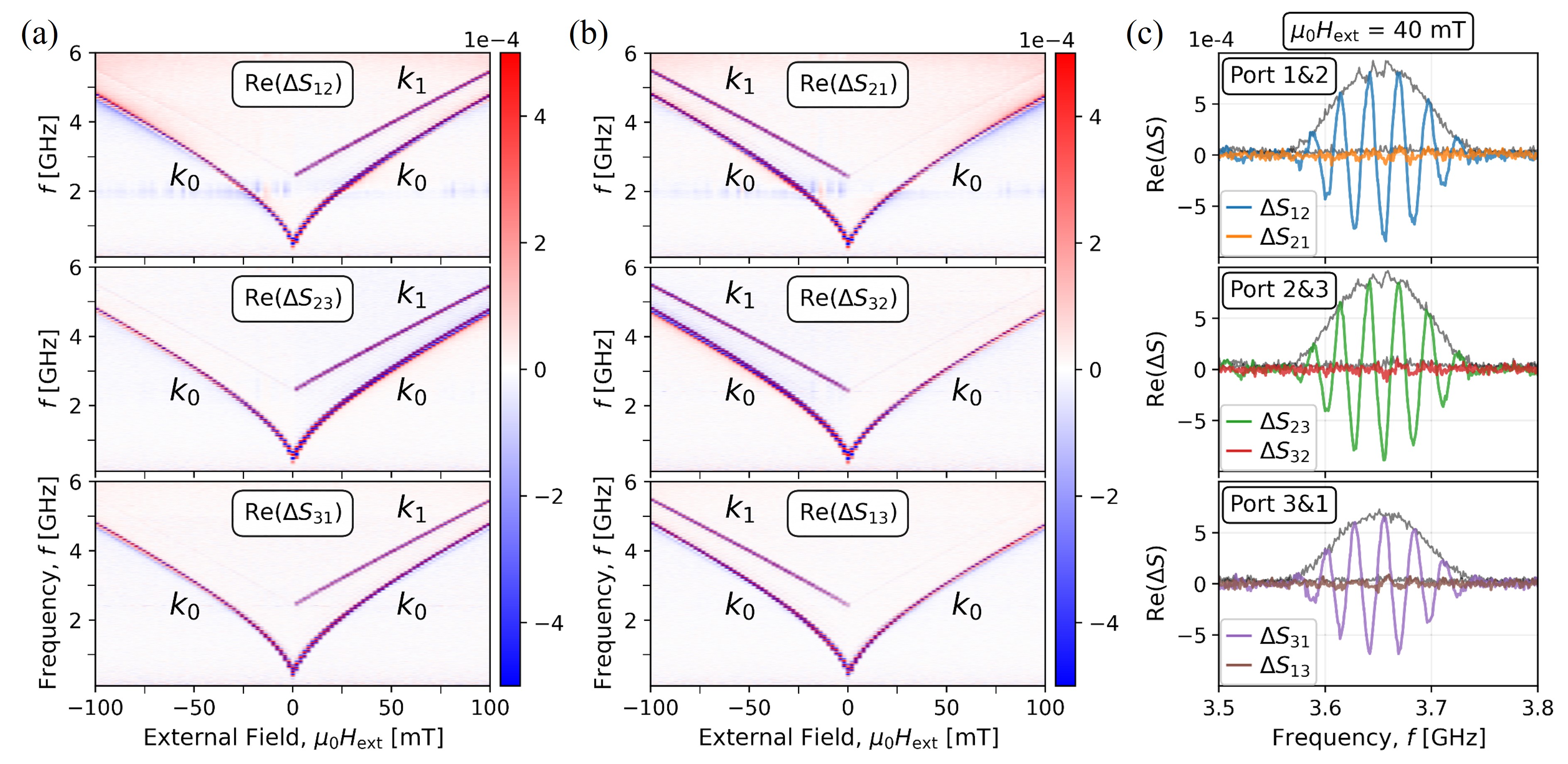}
    \caption{\label{fig:2 full_spectroscopy} 
    (a) Frequency-field mappings of the relative scattering parameters $\Delta S_{ij}$ for the three circulation channels. The $k_1$ branch shows clear amplitude asymmetry consistent with chiral excitation.
    (b) Complementary ($f,H_{\text{ext}}$) maps $\Delta S_{ji}$.
    (c) (color) Representative transmission spectra $\Delta \text{Re}(S_{ij})$ for $i,j \in \{1,2,3\}$ at $\mu_0 H_{\text{ext}}=40$\,mT, and their magnitude (grey). All data were acquired at an input power of $-15$\,dBm using a reference field of $\mu_0 H_{\text{ref}} = 200$\,mT.
    }
\end{figure*}

    The devices were characterized by 3-port spin-wave spectroscopy using a Rhode\&Schwarz-ZNA43 vector network analyzer \cite{RS_ZNA43}, and a PM8 probe station shown in Fig.~\ref{fig:1 Setup, SEM, CHirality, FFT}-(a) utilizing a combination of EDP-style dual (port 1 \& 2) and single (port 3) picoprobes \cite{GGB_dualprobes}. The whole setup was calibrated via a three-port SOLT (Short, Open, Load, Through) protocol using a CS-3 differential calibration substrate \cite{GGB_DiffCalSubstrate} in order to de-embed the magnonic signal from the rest of the setup.
    Figure ~\ref{fig:2 full_spectroscopy}-(a) displays the frequency-field map of the real part of $S_{12}$, $S_{23}$, $S_{31}$ in linear scale, namely the transmission from port $2$ to port $1$, port $3$ to port $2$, and port $1$ to port $3$, as indicated by the arrows in Fig.~\ref{fig:1 Setup, SEM, CHirality, FFT}-(b). Measurements are performed across an external field range from $-100$ to $+100$\,mT with a $2$\,mT increment, while the frequency was swept from $0.1$ to $6$\,GHz with $1601$ steps. For this mapping, we performed single spectral scans without averaging, while applying $-15$\,dBm of input power, and with an intermediate frequency bandwidth ($BW_{\text{IF}}$) of $500$\,Hz, which resulted in a $66$\,s acquisition time for each frequency sweep. 
    Owing to the weak inductive coupling, we chose to remove the parasitic baseline by performing a relative measurement of the scattering parameters $\Delta S = S_{\text{meas}} - S_{\text{ref}}$. The reference spectrum $S_{\text{ref}}$ was recorded at an external field $\mu_0 H_{\text{ref}}=200$\,mT sufficiently large to displace all spin-wave dynamics outside of the experimental frequency range. Additionally, in Appendix Fig.~\ref{fig:Appendix_Raw_data_and_power}-(a), we present typical raw transmission spectra across all ports at both low and high bias fields, where coherent interference between the propagating spin waves and direct electromagnetic leakage distorts the spectral envelope \cite{Kohl2026ModelingTransducers}. Besides, one notices a much higher baseline between port 1 and 2 mostly due to the configuration of the dual picoprobes, which inevitably leads to higher electromagnetic crosstalk than with the single probe configuration. For these reasons, we prefer to represent the microwave transmission through spin waves as relative measurements $\Delta S$.
    The transmission spectra on this device reveal two branches corresponding respectively to the first two peaks $k_0 \approx 0$ and $k_1=\frac{2\pi}{a}\approx12.6$\,rad/$\mu$m, for which we expect the most efficient spin-wave generation to occur from the excitation field's spectral density shown in Fig.~\ref{fig:1 Setup, SEM, CHirality, FFT}-(d). The latter branch ($k_1$) possesses a strong amplitude asymmetry between positive and negative fields, consistently with the cross-product rule of Eq.(\ref{eq:RightHandRule}), while the $k_0$ branch remains largely symmetric. This difference in asymmetry between the $k_0$ and $k_1$ branches is expected from a more pronounced circular ellipticity of the excitation field felt by the higher wavevectors, which is a requirement for the chiral coupling \cite{Yu2021ChiralRadiation,Temdie2023ChiralGrating, Wagle2026ShapingBeams}. The reciprocal spectra $S_{21}$, $S_{32}$, $S_{13}$ shown Fig.~\ref{fig:2 full_spectroscopy}-(b) display identical spin-wave features with opposite applied field polarity, consistent with Eq.(\ref{eq:RightHandRule}). Additionally, Figure ~\ref{fig:2 full_spectroscopy}-(c) shows typical spectra at $\mu_0 H_{\text{ext}}=40$\,mT of all transmission S-parameters, namely both $\Delta S_{ij}$ and $\Delta S_{ji}$ for $i,j \in \{1,2,3\}$ and $i \neq j$, confirming the unidirectional nature of the beams excited by the grating (see also Appendix Fig.~\ref{fig:Appendix_Example_Spectra_DeltaL} for representative spectra over $400$\,mT field range). These distinct contrasts between passing and isolated ports demonstrates a genuine circulation of spin-waves between all three ports.\\
\begin{figure*}[!htbp]
    \includegraphics[width=\textwidth]{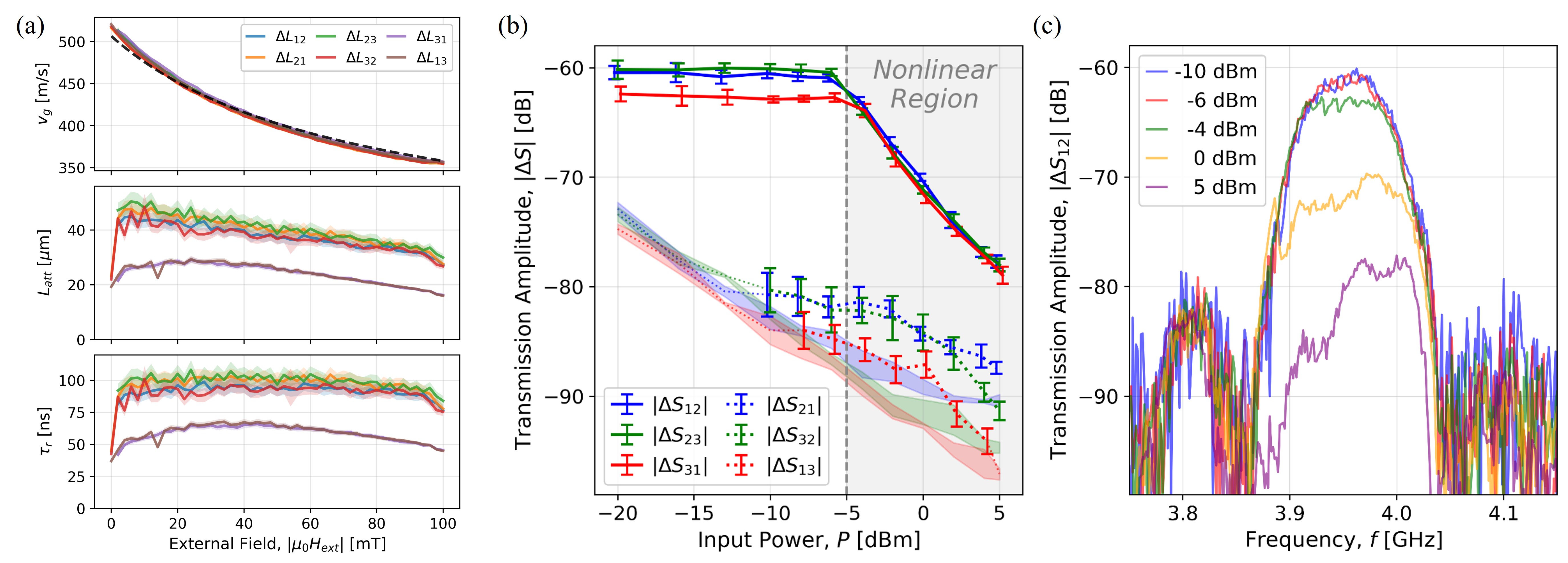}
    \caption{
            (a) Bias field dependence of the spin-waves transport properties derived from the inductances $\Delta L_{ij}$ (extracted from the data in Fig.~\ref{fig:2 full_spectroscopy}) in the passing polarity of the bias field. (top) group velocity $v_g$, where the dashed line represents the theoretical calculation based on the dispersion relation Eq.(\ref{disp_ip}), (middle) attenuation length $L_{\text{att}}$, and (bottom) relaxation time $\tau_{\text{r}}=\frac{L_{\text{att}}}{v_g}$. Shaded bands denote the estimated uncertainties for $L_{\text{att}}$ and $\tau_{\text{r}}$, whereas uncertainties for $v_g$ fall within the linewidth.
            (b) Power dependence of the transmission amplitudes of the $k_1$ peaks at $\mu_0 H_{\text{ext}}=50$\,mT in the passing (solid line), and isolated (dashed line) direction, including an estimated noise floor level (shaded bands). 
            (c) Spectra comparison showing the spectral distortion and amplitude saturation of $\Delta S_{12}$ past the non-linear threshold at $-5$\,dBm. The measurements in (b) and (c) were acquired using a reference field of $\mu_0 H_{\text{ref}} = 300$\,mT.}
    \label{fig:3 Power Comparison}
\end{figure*}

    By converting our measured $3 \times 3$ scattering matrices into impedance matrices (as detailed in Appendix~\ref{Appendix Example Spectra}), we obtain the dynamic inductance spectra $\Delta L_{ij}$. From these, we summarize in Figure~\ref{fig:3 Power Comparison}-(a) the field dependence of the extracted spin-wave propagation properties \cite{Loayza2018}, which include the group velocity $v_g$, the attenuation length $L_{\text{att}}$, and the corresponding relaxation time $\tau_{\text{r}} = L_{\text{att}}/v_g$. These characteristics demonstrate that the prototype functions effectively as a combined non-reciprocal circulator and programmable delay line. This field-dependent delay provides a powerful tool to directly investigate spin-wave propagation dynamics. Experimentally, the group velocity was extracted from time-of-flight measurements via $v_g = \frac{d\omega}{dk} = \frac{d}{\tau_{\text{ToF}}}$, where $\tau_{\text{ToF}} = -\frac{d\phi(\omega)}{d\omega}$ and $\phi(\omega)$ is the phase of the transmission spectra $\Delta L_{ij}$. It reveals an identical monotonic decrease across all transmission channels as the bias field increases, from $515$\,m.s$^{-1}$ at $0$\,mT to $355$\,m.s$^{-1}$ at $100$\,mT (with a mean fit uncertainty of $\sigma_{v_g} \approx 1.1$\,m.s$^{-1}$), in very good agreement with the theoretically calculated group velocity (indicated by dashed lines), derived from the dispersion relation \cite{Kalinikos1986TheoryConditions}:
        \begin{equation}
            \omega_{\text{res}}^2(k,H) = \Omega_k (\Omega_k + \omega_M) + \omega_M^2 P(1-P)
            \label{disp_ip}
        \end{equation}
    where $\omega_M=\gamma \mu_0 M_s$, $\Omega_k= \gamma \mu_0 H_{\text{equ}} +{\omega_M}\,\Lambda^2 k^2$ with $\Lambda^2=\dfrac{2A_x}{\mu_0M_s^2}$ the exchange length, and $P=1-\frac{1-e^{-kt}}{kt}$, which reflects the influence of the dynamic dipolar interaction.    
    To estimate the attenuation length $L_{\text{att}}$, we consider an exponential decay and estimate the ratio of the mutual inductance amplitude to the absorptive part of the self-inductance $\left| L_{ij} / L_{jj} \right| \approx \exp(-\frac{d}{L_{\text{att}}})$ (see also Appendix Fig.~\ref{fig:Appendix_L_Ratio}), where $d$ is the inter-transducer distance. This approach, which effectively de-embeds the local transduction efficiency, shows a mild decrease of $L_{\text{att}}$ with the bias field, yielding similar values for the channels $1$--$2$ and $2$--$3$ ranging from $60$\,$\mu$m at low fields down to $40$\,$\mu$m at $100$\,mT (with a mean standard error of $\sigma_{L_{\text{att}}} \approx 2.3$\,$\mu$m), while the channel $3$--$1$ exhibits slightly reduced attenuation length ranging from $30$\,$\mu$m to $20$\,$\mu$m ($\sigma_{L_{\text{att}}} \approx 0.8$\,$\mu$m) respectively. This slight reduction, which is also visible in the amplitude of the transmission spectra, is likely due to the shape of our prototype which requires a longer termination of port $1$ in order to couple with port $3$, resulting in additional ohmic loss compared with the two other ports. This minor weakness could be corrected with a proper microwave engineering of the scheme in order to balance all ports. Consequently, the calculated relaxation time $\tau_{\text{r}} = L_{\text{att}}/v_g$ remains remarkably stable across the entire field range, averaging near $100$\,ns ($\sigma_{\tau_{\text{r}}} \approx 5.6$\,ns) for channels $1$--$2$ and $2$--$3$, and $\approx 60$\,ns ($\sigma_{\tau_{\text{r}}} \approx 2.0$\,ns) for channel $3$--$1$. These results demonstrate that the chosen inter-transducer distance, $d = 15\ \mu\text{m}$, is well within the attenuation length across the entire field range, confirming that the observed insertion loss is not primarily limited by spin-wave decay, but is instead dominated by the poor impedance radiation matching at this length scale \cite{Connelly2021,Vanderveken2022, Bruckner2025MicromagneticTransducers}.\\ 
    The performance and power handling of the circulator were separately investigated for various input powers $P_{\text{in}}$ ranging from $-20$\,dBm to $5$\,dBm, and various bias field up to $200$\,mT. Figure~\ref{fig:3 Power Comparison}-(b) depicts the power dependence maximum transmission amplitudes for the $k_1$ branch in both the passing and the blocking directions across all ports at a bias field of $\mu_0 H_{\text{ext}}=50$\,mT. Below $P_{\text{in}} = -6$\,dBm, the device behaves linearly with the relative spin-wave transmission magnitude, independent of the input power, pointing around $-60$\,dB. It shows a difference of more than $20$\,dB between the passing and the blocking directions, with backward signals remaining indistinguishable from the experimental noise floor indicated by the dotted line in the shaded regions. As the power exceeds $-6$\,dBm, the system enters a non-linear regime characterized by spectral distortion and amplitude reduction. These findings are consistent with the non-linear power-limiting effects observed by Davidkova et al.\cite{Davidkova2025NanoscaleTechnology}, which is due to the onset of multi-magnon scattering processes in YIG thin films. This transition is resolved in Fig.~\ref{fig:3 Power Comparison}-(c), which superimposes the forward transmission spectra $|\Delta S_{12}|$ at varying power levels. While the response remains consistent up to $-6$\,dBm, higher powers induce progressive peak compression and broadening, altering the operation of the circulator.
    Notably, the non-linear threshold varies slightly with the bias field, shifting from approximately $-5$\,dBm at $50$\,mT to $-6$\,dBm at $100$\,mT, before increasing to $-3$\,dBm at $200$\,mT (as shown in Appendix  Fig.~\ref{fig:Appendix_Raw_data_and_power}-(b)). The observed increase in the non-linear threshold qualitatively aligns with established theories on high-power microwave instability \cite{Suhl1955}. However, because our prototype's return loss varies across the frequency range, this non-uniform return loss slightly disrupts the expected monotonous increase of the non-linear threshold.\\

    To further explore the limits of frequency selectivity, and the potential for multi-band operation in magnonic circuits, we fabricated a second device with a reduced width $w=100$\,nm, periodicity $a=400$\,nm, and an inter-transducer distance $d=10$\,$\mu$m (refer to inset of Fig.~\ref{fig:4 Device 2}-(a)).
    The discrete wavevectors allowed by the grating periodicity, occurring at $k_n = n \cdot 2\pi/a$, have their relative excitation strength sampled by the single-bar envelope as indicated by the red dashed line in Fig.~\ref{fig:4 Device 2}-(a), and whose first zero is located at $2\pi/w \approx 62.8$\,rad/$\mu$m. By optimizing the $a/w$ ratio to a value of $4$ (an increase from the $2.5$ ratio used in our first device), we effectively enhance the excitation strength of this next wavevector $k_2 \approx 31.4\,\text{rad}/\mu\text{m}$. Physically, this geometric ratio dictates how many wavevector peaks of the excitation comb fit within the main lobe of the single-bar envelope, thereby increasing the spectral weight at $k_2$.\\
    
    The experimental validation of this multi-wavevector excitation is captured in the frequency-field mapping of the transmission magnitude $\Delta S_{31}$ shown in Fig.~\ref{fig:4 Device 2}-(b) in a bias field range up to $50$\,mT. The map clearly resolves three distinct branches: a dominant and symmetric $k_0 \approx 0$ branch, alongside then non-reciprocal $k_1$ and $k_2$ branches. Additionally, intermediate small ripple structures, also exhibiting similar non reciprocity, are visible between $k_0$ and $k_1$, and correspond to the secondary lobes of the grating's Fourier transform.
    We show in Figure~\ref{fig:4 Device 2}-(c) a typical spectra at $\mu_0 H_{\text{ext}}=20$\,mT for this device zoomed around the peaks corresponding to the wavevector $k_1$ and $k_2$. Despite the inherent increase in insertion losses at higher wavevectors, reaching $67$\,dB for $k_1$ and $85$\,dB for $k_2$, both remain non-reciprocal, with a chiral isolation of $23$\,dB for $k_1$ and at least $14$\,dB for $k_2$ above the noise floor. The residual signal observed in the isolation direction of $k_1$ is likely attributed to a non-complete chiral coupling, or perhaps a parasitic absorption and re-emission at the intermediate port $2$.\\
    Finally, the frequency dependence of the propagation properties, quantified through time of flight measurements $\tau_{\text{ToF}}$ and group velocity $v_g = d/\tau_{\text{ToF}}$, is shown in Fig.~\ref{fig:4 Device 2}-(d). We observe a distinct field-dependent behavior between the two branches. For the $k_1$ mode, $\tau_{\text{ToF}}$ increases from $19$\,ns to $23$\,ns (with an average standard error of $\sigma_{\tau_{\text{ToF}},k_1} \approx 0.30$\,ns) as the bias field is swept up to $50$\,mT, corresponding to a group velocity comparable to the one extracted for the first device. In stark contrast, the $k_2$ branch maintains a remarkably invariant time of flight of $\tau_{\text{ToF}} \approx 16$\,ns (with $\sigma_{\tau_{\text{ToF}},k_2} \approx 0.12$\,ns) across the same field range, which results in a constant group velocity of $\approx 635$\,m.s$^{-1}$ greater than the one for $k_1$ (with average uncertainties of $\sigma_{v_g,k_1} \approx 6.5$\,m.s$^{-1}$ and $\sigma_{v_g,k_2} \approx 4.9$\,m.s$^{-1}$).
    This disparity is fundamentally rooted in the underlying dispersion relation Eq.(\ref{disp_ip}). Namely, the influence of the dynamic dipolar term in the dispersion is more pronounced at the lower wavevector $k_1$, making its curvature more sensitive to the Zeeman shift of the spin-wave manifold. However, at the higher wavevector $k_2$, the dispersion enters a regime increasingly dominated by the quadratic exchange term ($\propto k^2$). In this exchange-active region, the dispersion slope becomes less affected by external field variations, resulting in the observed constant group velocity. This transition provides a versatile platform offering different timing characteristics for signal routing in miniaturized microwave architectures, where both field-tunable and field-robust time-delays can be exploited simultaneously.\\

    In summary, we have successfully implemented a comprehensive three-port spin-wave spectroscopy to characterize the circulation of microwave power at a micron-scale.
    This experimental platform allows for the de-embedding of multi-channel scattering parameters, providing a rigorous verification of non-reciprocal signal routing in integrated magnonic circuits. We showcase a magnonic circulator prototype based on 3 pairs of nano-gratings, acting as a three rectilinear beam channel design operating over a working area of $100\times 100$\,$\mu$m$^2$. This device, a first of its kind, exhibits quasi even level of transmission amplitude, and isolation greater than $20$\,dB between ports, allowing for a proper microwave circulation. Further engineering of the grating’s dimensions enables the excitation of additional higher-order wavevector branches, offering a versatile route for the development of multi-band and frequency-selective magnonic components. These findings establish a scalable pathway for the integration of non-reciprocal technologies into high-frequency 5G/6G architectures and cryogenic quantum systems.\\\\

\begin{acknowledgments}
    This project has received co-funding from the European Union’s Horizon Europe Research and Innovation Program under Grant Agreement No. 101126644 SEED-HORIZON-MSCA-2022-COFUND-01.
    It further benefited from a government grant operated by the French National Research Agency as part of the France 2030 program, Reference No. ANR-22-EXSP-0004(SWING), as well as the ANR project MagFunc/ANR-20-CE91-0005. M.L. received funding from the German Bundesministerium für Wirtschaft und Energie (BMWI) under Grant No. 49MF180119.
    We thank Hicham Majjad, Thibaut Devolder, and Asma Mouhoub for their help with the nanofabrication and Mufti Avicena for meaningful discussions. 
\end{acknowledgments}

\appendix
\section{Raw Data and Power} \label{Appendix Raw Data}
\begin{figure*}[!htbp]
    \centering
    \begin{tabular}{cc}
      \includegraphics[width=0.69\textwidth]{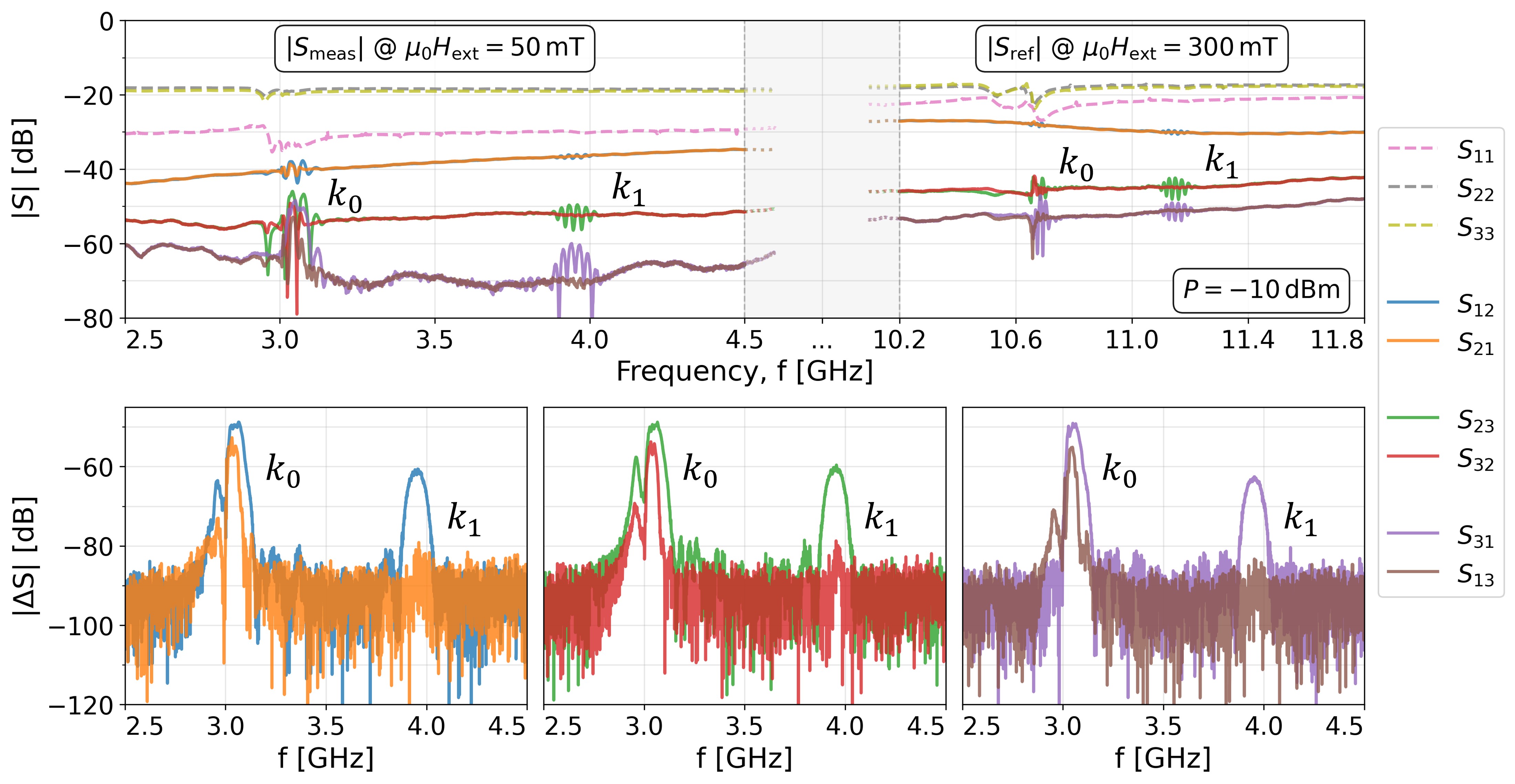} &
      \includegraphics[width=0.30\textwidth]{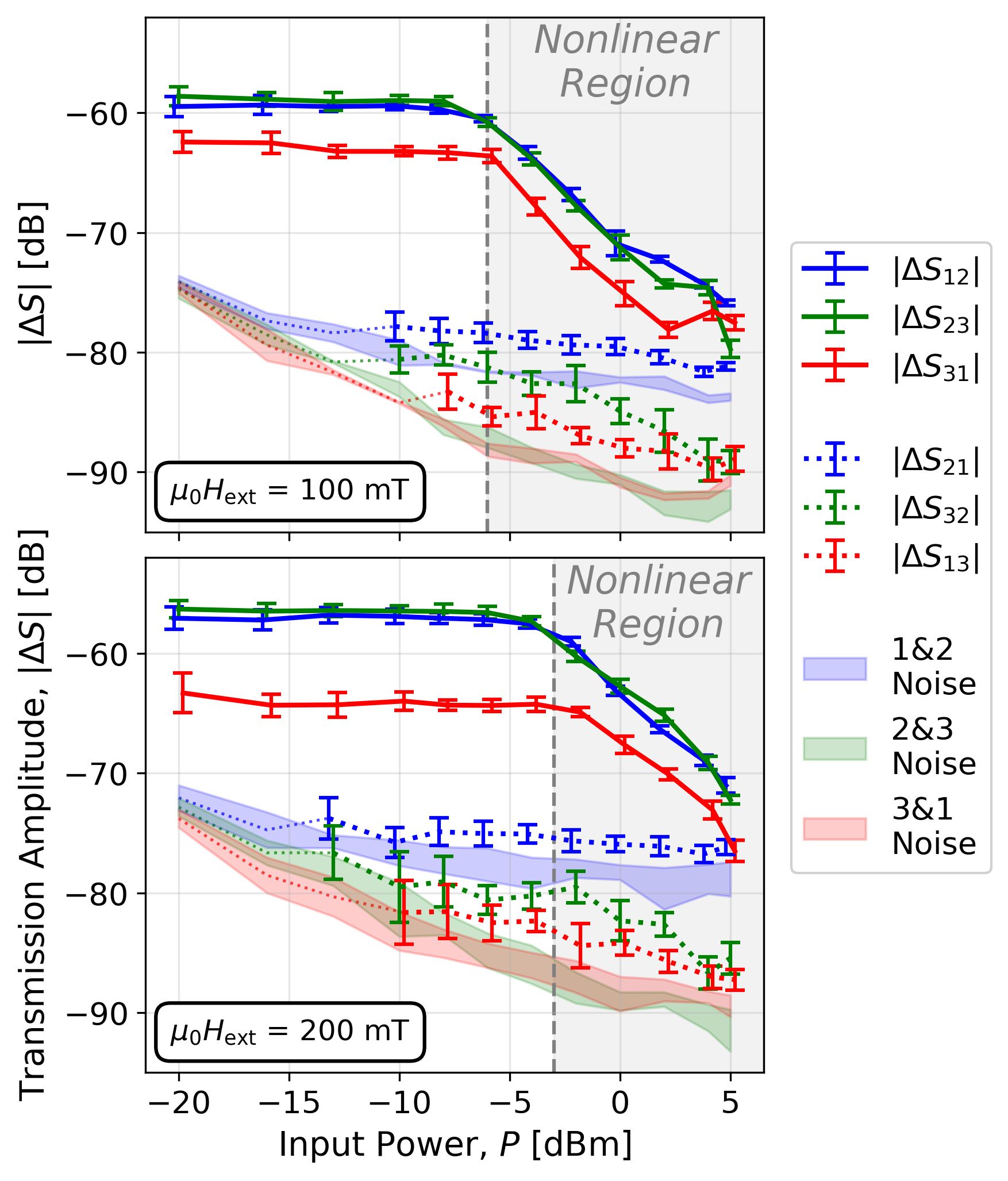}\\
      (a) & (b)
    \end{tabular}
    \caption{
        (a) Raw data with insertion and rejection loss. 
        (b) Power dependence at $\mu_0 H_{ext} = \{100$\,mT, $200$\,mT$\}$ at $\mu_0 H_{ref}=300$\,mT.
    }        
    \label{fig:Appendix_Raw_data_and_power}
\end{figure*}

    The background-subtraction procedure used to isolate the magnonic signal from electromagnetic (EM) crosstalk is illustrated in Fig.~\ref{fig:Appendix_Raw_data_and_power}-(a). The upper panel displays the raw transmission magnitude $|S_{ij}|$ at $\mu_0 H_{\text{ext}}=50$\,mT compared to a reference baseline recorded at $\mu_0 H_{\text{ext}}=300$\,mT (input power $P=-10$\,dBm). These raw spectra reveal that the inductive crosstalk is not uniformly balanced across the device, namely the EM coupling is most pronounced between ports $1$ and $2$ with a base line at $-40$\,dB, while the port pair $3$--$1$ exhibits the lowest background levels at $-65$\,dB. This differences in the crosstalk between the dual and single picoprobes is a limitation inherent to the shape of this EDP-style GGB dual-picoprobes, which unfortunately don't allow to tackle small signal below this baseline.
    The baseline-corrected signals (lower panel), defined as $\Delta S_{ij} = S_{ij}(H) - S_{ij}(H_{\text{ref}})$, clearly resolve the underlying magnonic resonances. At this specific power level ($-10$\,dBm), the low-$k$ branch ($k_0$) is already significantly distorted, indicating operation within the non-linear regime. In contrast, the $k_1$ peak retains a well-defined Gaussian lineshape and possess high directional contrast, suggesting a higher threshold for non-linear instabilities at larger wavevectors, in accordance with the wavevector dependent excitation efficiency given by the spectral density of the excitation field illustrated in Fig.~\ref{fig:1 Setup, SEM, CHirality, FFT}-(d) and Fig.~\ref{fig:4 Device 2}-(a).
    Figure~\ref{fig:Appendix_Raw_data_and_power}-(b) presents the power-dependent analysis for external fields of $100$\,mT and $200$\,mT. While the qualitative evolution of the spectra matches the $50$\,mT data discussed in the main text, the non-linear power threshold—marked by the onset of peak compression and broadening—varies with the bias field. We observe a minor reduction of this threshold to $-6$\,dBm at $100$\,mT, and an increase to $-3$\,dBm at $200$\,mT, which is somewhat consistent with the known $3/4$ power law of magnetic instability at high microwave power \cite{Suhl1955}. For quantitative comparison within the linear regime, the corresponding forward-to-reverse isolation ratios $C_{ij}$ evaluated at $-6$\,dBm across all three bias fields are summarized in the following Table~\ref{tab:Appendix_isolation_ratios}.

\begin{table}[h]
    \centering
    \caption{\label{tab:Appendix_isolation_ratios}Forward-to-reverse transmission isolation ratios $C_{ij} = 20 \log_{10} |S_{ji} / S_{ij}|$ and corresponding resonance frequencies $f_{\text{res}}$ of the $k_1$ wavevector branch for Device~\#1 at an excitation power of $-6\,\text{dBm}$ under various external bias fields (corresponding to Fig.~\ref{fig:3 Power Comparison}(b) and Appendix Fig.~\ref{fig:Appendix_Raw_data_and_power}(b)).}
    \label{tab:isolation_ratios}
    \begin{ruledtabular}
    \begin{tabular}{ccccc}
        $\mu_0 H_{\text{ext}}$ & $f_{\text{res}}$ & $C_{12}$ & $C_{23}$ & $C_{31}$ \\
        (\text{mT}) & (\text{GHz}) & (\text{dB}) & (\text{dB}) & (\text{dB}) \\
        \colrule
        50  & 3.96  & $20.92 \pm 1.28$ & $21.68 \pm 2.40$ & $22.08 \pm 1.71$ \\
        100 & 5.42  & $17.87 \pm 1.05$ & $20.48 \pm 1.60$ & $21.77 \pm 1.31$ \\
        200 & 8.32  & $17.87 \pm 1.54$ & $24.00 \pm 1.70$ & $18.14 \pm 2.01$ \\
    \end{tabular}
    \end{ruledtabular}
\end{table}

\section{Multi-band Circulation and Maps} \label{Complementary Map}
\begin{figure*}[!htbp]
    \centering
    \begin{tabular}{cc}
      \includegraphics[width=0.38\textwidth]{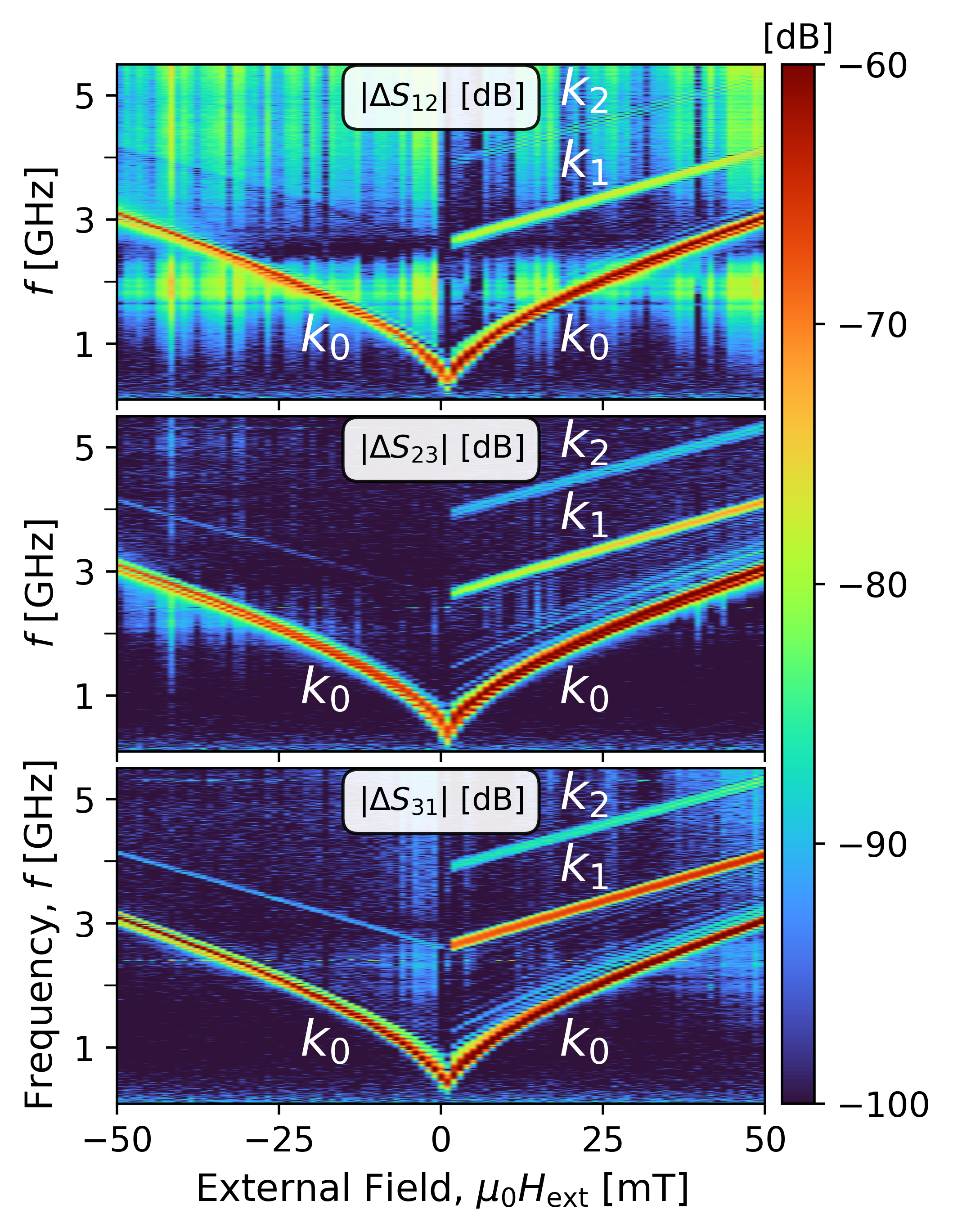} &
      \includegraphics[width=0.38\textwidth]{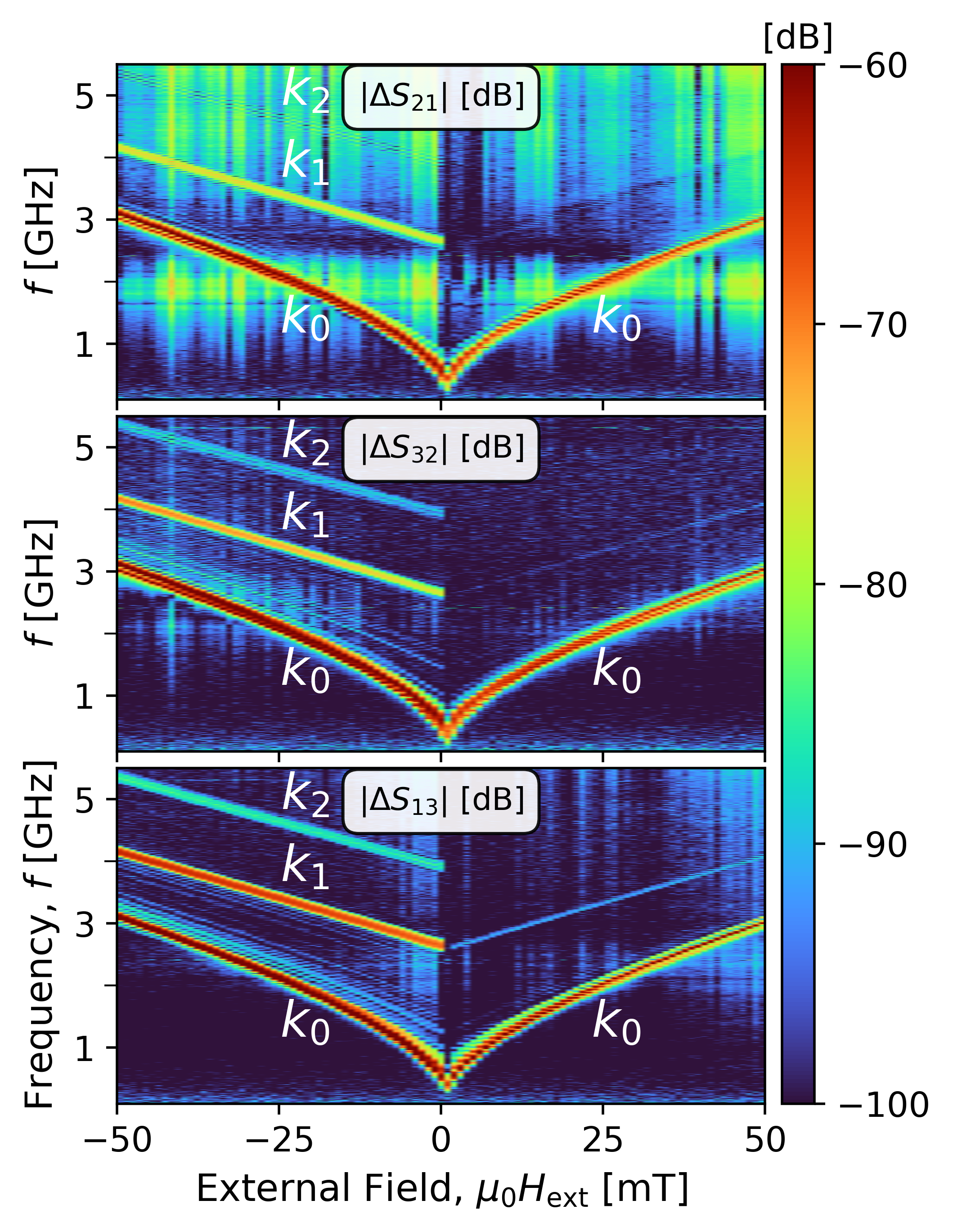}\\
      (a) & (b)
    \end{tabular}
    \caption{
            (a) Full ($f, H_{\text{ext}}$) mapping and
            (b) complementary transmission maps for the second device. Data were acquired using a reference field of $\mu_0 H_{\text{ref}} = 250$\,mT at an input power of $-5$\,dBm.}
    \label{fig:Appendix_Complementary_Maps}
\end{figure*}

    The comprehensive 3-port characterization of the second investigated device ($w=100$\,nm, $a=400$\,nm) is shown in the complete set of frequency--field  mappings in Figure~\ref{fig:Appendix_Complementary_Maps}-(a,b), resolving both the $k_1$ and $k_2$ branches across all ports. This second data set highlights the reproducibility of the chiral excitation mechanism and demonstrates the feasibility of higher-order wavevector engineering for multi-band magnonic applications.\\
    
\section{$S$-Parameter Conversion and Inductance Spectra} \label{Appendix Example Spectra}
\begin{figure*}[!htbp]
    \includegraphics[width=\textwidth]{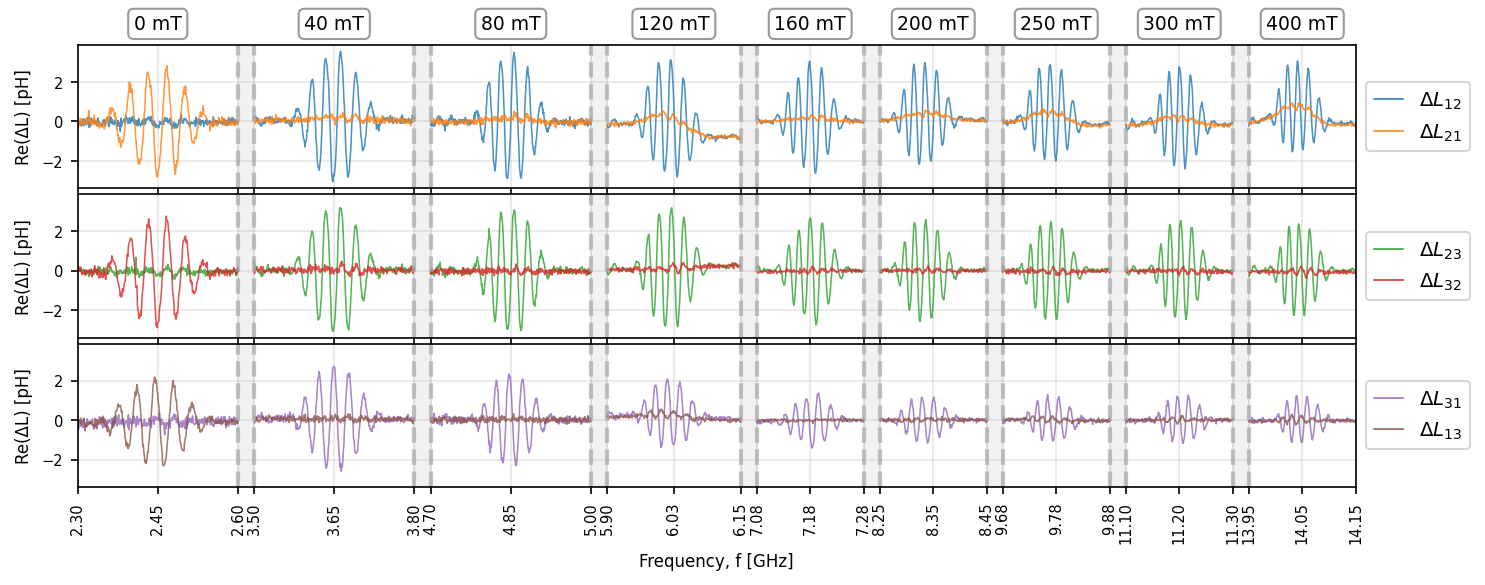}
    \caption{Example spectra $\Delta L_{ij}= L_{ij}(H_{\text{ext}}) - L_{ij}(H_{\text{ref}})$ of the $k_1$ branch for the first device, measured at an input power of $-10$\,dBm. Data are shown for external bias fields $\mu_0 H_{\text{ext}} = 0$, $40$, $80$, and $120$\,mT using a reference field of $\mu_0 H_{\text{ref}} = 300$\,mT, and for $\mu_0 H_{\text{ext}} = 160$, $200$, $250$, $300$, and $400$\,mT using $\mu_0 H_{\text{ref}} = 10$\,mT.}
    \label{fig:Appendix_Example_Spectra_DeltaL}
\end{figure*}

    We display in Figure~\ref{fig:Appendix_Example_Spectra_DeltaL} a panel of inductance spectra over the full $400$\,mT range of bias field. To extract these spectra, the measured $3 \times 3$ scattering matrix $\mathbf{S}$ must be transformed into the impedance matrix $\mathbf{Z}$. For a general multi-port network, this transformation is defined as:$$\mathbf{Z} = Z_0 (\mathbf{I} + \mathbf{S})(\mathbf{I} - \mathbf{S})^{-1}$$where $\mathbf{I}$ is the $3 \times 3$ identity matrix and $Z_0 = 50\,\Omega$ is the characteristic impedance of the measurement system. From the resulting impedance matrix, the frequency-dependent complex inductance is defined as:$$L_{ij} = \frac{Z_{ij}}{2\pi \mathrm{i} f}$$Applying the reference field subtraction yields the purely dynamic magnonic inductance $\Delta L_{ij} = L_{ij}(H_{\text{ext}}) - L_{ij}(H_{\text{ref}})$.
    Figure~\ref{fig:Appendix_Example_Spectra_DeltaL} shows that the amplitude of the spin-wave transmission signal remains fairly constant for the channel $12$, and $23$, while it decreases slightly for the channel $31$.

\section{Inductance Ratio} \label{Appendix Inductance Ratio}
\begin{figure*}[!htbp]
    \includegraphics[width=\textwidth]{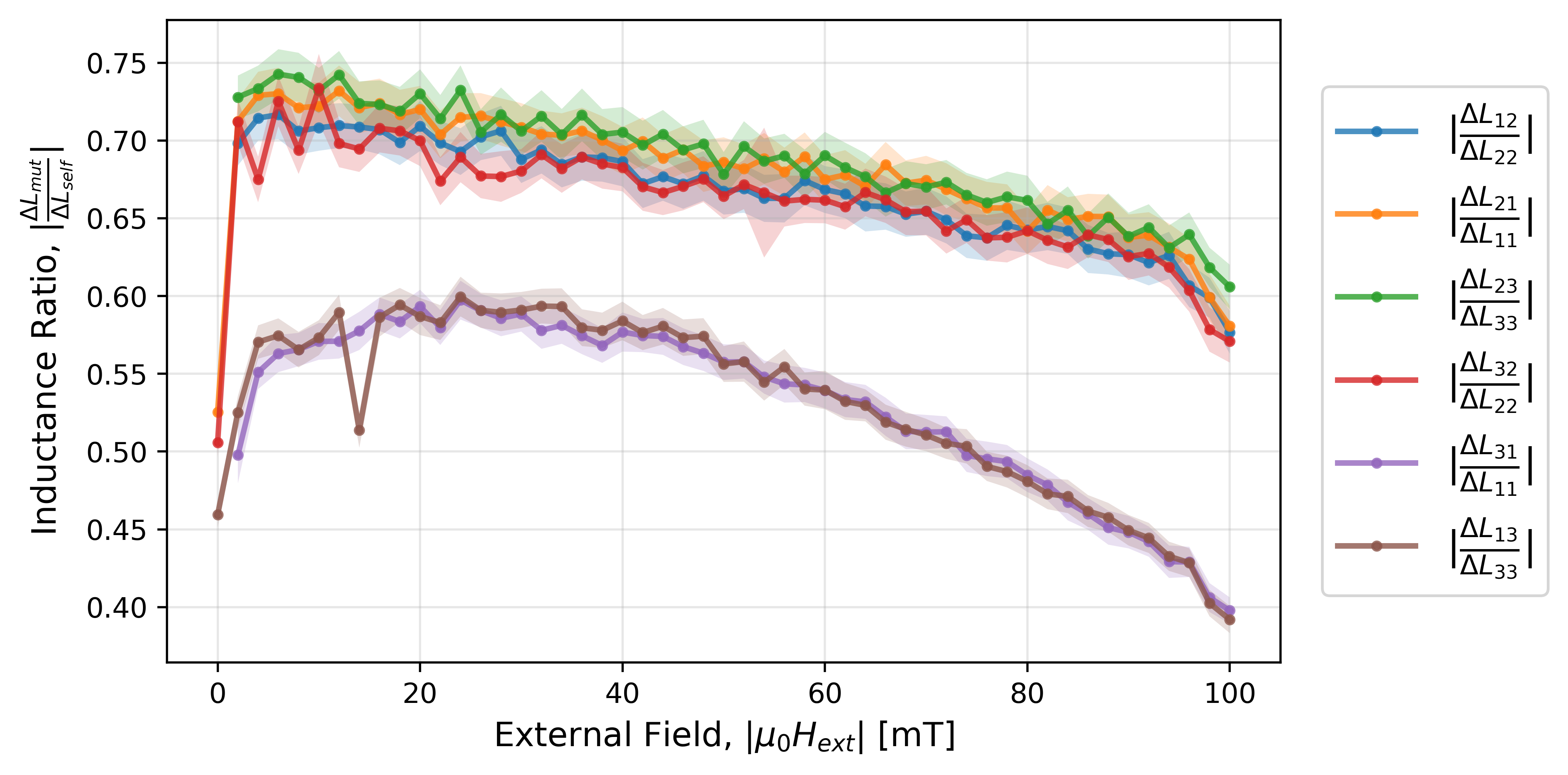}
    \caption{Inductance amplitude ratios for attenuation length estimation for the first device at. The ratio of the mutual (transmission) inductance amplitude to the absorptive part of the self- (reflection) inductance, $|L_{ij} / L_{jj}|$, is plotted as a function of the external magnetic field magnitude $|H_{\text{ext}}|$, corresponding to the passing polarity of the bias field for each respective channel. Shaded bands denote the estimated fitting errors.}
    \label{fig:Appendix_L_Ratio}
\end{figure*}

    To further elucidate the extraction of the attenuation length $L_{\text{att}}$, Figure~\ref{fig:Appendix_L_Ratio} presents the field-dependent evolution of the mutual-to-self-inductance ratio, $|L_{ij} / L_{jj}|$. This metric effectively isolates the intrinsic spin-wave propagation dynamics by normalizing out local transduction efficiencies. As shown, propagation channels $1$--$2$ and $2$--$3$ exhibit highly symmetric behavior, with their inductance ratios mildly decreasing from approximately $0.8$ at low fields down to $0.6$ (with mean standard error $\sigma_{\text{ratio}} \approx 0.015$) at $100$\,mT. In contrast, channel $3$--$1$ yields systematically lower values, decreasing from roughly $0.6$ down to $0.4$ ($\sigma_{\text{ratio}} \approx 0.011$) over the same field range. This disparity mirrors the slightly reduced attenuation length and lower transmission amplitude observed for this channel in the main text.

\bibliography{ReferencesLinCirc}

\end{document}